\documentclass [12pt]{article}
\usepackage {graphicx}
\usepackage {longtable}

\begin{document}
\begin{center}
\section*{The Cosmic Organism Theory}
\end{center}

\bigskip

\begin{center}
Ding-Yu CHUNG* and Volodymyr KRASNOHOLOVETS**
\end{center}

\bigskip

\begin{center}
* \textit{P.O. Box 180661, Utica, Michigan 48318, U.S.A. \break e-mail:
chung@wayne.edu}
\end{center}

\begin{center}
\textit{**Institute for Basic Research, 90 East Winds Court, Palm Harbor, FL
34683, U.S.A., e-mail: v\_kras@ayahoo.com}
\end{center}

\bigskip

\small{
\textbf{Abstract.} We present the cosmic organism theory in which all
visible and invisible matter has different cosmic genetic expressions. The
cosmic gene includes codes for the object structure and the space structure.
The cosmic digital code for the object structure consists of full object (1,
2, and 3 for particle, string, and membrane, respectively) and empty object
(0) as anti de Sitter space (AdS). The tessellation lattice of empty objects
is tessellattice. The decomposition of a full object in tessellattice
results in the AdS/CFT (conformal field theory) duality. The digital code
for the object structure accounts for the AdS/CFT duality, the dS/bulk
duality, and gravity. The digital code for the space structure consists of 1
and 0 for attachment space and detachment space, respectively. Attachment
space attaches to object permanently at zero speed or reversibly at the
speed of light. Detachment space detaches from the object irreversibly at
the speed of light. The combination of attachment space and detachment space
results in miscible space, binary lattice space or binary partition space.
Miscible space represents special relativity. Binary lattice space consists
of multiple quantized units of attachment space separated from one another
by detachment space. Binary lattice space corresponds to the nilpotent
universal computational rewrite system (NUCRS) by Diaz and Rowlands. The
gauge force fields and wavefunction are in binary lattice space. Binary
partition space consists of separated continuous phases of attachment space
and detachment space. With tessellattice and binary lattice space, 11D brane
is reducing to 4D particle surrounded by gravity and the gauge force fields.
The cosmic dimension varies due to different speeds of light in different
dimensional space-times and the increase of mass. The force fields in
tessellattice are the original gravity before the big bang and the strong
force in nucleus. The cosmology model and dark energy are predicted by the
Santilli isodual theory and the Dirac hole theory. }

\subsection*{1. Introduction}

In the cosmic organism theory (Chung, 2002a), different universes are
different cosmic organs for the cosmic organism that is the multiverse.
Different universes are the different expressions of the common cosmic gene.
The cosmic gene is the cosmic digital code (Chung, 2002b) in the cosmic
dimension. The cosmic digital code includes the codes for the object
structure and the space structure.

In the analysis of the multiverse (Ellis, Kirchner, and Stoeger (2004) state
that there is a definite causal connection, or ``law of laws'', relating all
the universes in these multiverses. Law of laws can be described as the
ultimate law of fundamental laws, which include relativity, quantum
mechanics, and the laws for the existence of different physical constants,
dimensionality, particle content, and the size of universe in the
multiverse. The ultimate law connects fundamental laws. In the cosmic
organism theory, the ultimate law is the cosmic gene. The cosmic organism
theory follows the Alfred North Whitehead's philosophy of organism.
According to Whitehead (Whitehead, 1929), the actual world is a process, and
the process is the becoming of actual entities. An actual entity is not an
inert and permanent substance, but a relational process of becoming. Its
`being' is constituted by its `becoming'. Michel Bounias (2002) applied the
Hamiltonian concept to living organism in the evolutionary process.

The paper is divided into seven sections: the introduction, the cosmic
digital code for the object structure, the digital code for the space
structure, the cosmic dimension, cosmology, force fields, and the summary.

\subsection*{2 The digital cosmic code for the object structure}

In terms of the current hypothesis, which is still shared by the majority of
physicists, fundamental principles of the world are associated with
so-called strings and membranes - abstract entities that are treated as
primary elements of the world. Then the cosmic digital code for object
structure can be presented by the full object, which involves particle,
string and membrane (1, 2, and 3, respectively), and the empty object (0) as
anti de Sitter space (AdS). Full object occupies space fully, while empty
object is originally as the gap among full objects.

An empty cell (or topological ball, or superparticle), as a primary element,
was first proposed by Bounias and Krasnoholovets (2003a,b) in the
tessellattice theory. A tessellation lattice of empty cells with the size of
the Planck is the tessellattice. A particle is treated as a local
deformation of the tessellattice. The behavior of a moving particle in a
cellular space studied by Krasnoholovets (1993,1997,2002) is characterized
by a submicroscopic mechanics that discloses the particle's hidden dynamics
behind the formalism of conventional quantum mechanics. It is important that
in submicroscopic mechanics a moving particle interacts with coming cells of
the space. Considering space in terms of the tessellattice, Bounias and
Krasnoholovets (2003a,b,c) concluded that a moving canonical particle
experiences a fractal decomposition into a core particle surrounded by its
inerton cloud (the notion of an \textit{inerton} as a carrier of inert
properties of the particle, which thus becomes a substrructure of the
particle's matter waves, was introduced in works by Krasnoholovets and
Ivanovsky (1993) and Krasnoholovets (1997)). The decomposition generates the
gravitational mass, and inerton carries the inertial of gravitational mass
of particles. Gravity comes from the continuous generation of inerton clouds
by the continuous interaction of a moving particle and the empty cells in
the tessellattice.

Empty object as AdS is the active empty cell that can decompose a full
object. In AdS, the cosmological constant ($\Lambda ) < 0$, indicating a
negative expansion. Such negative expansion can reduce space dimension from
non-zero to zero. Empty object as AdS, therefore, can detach one of the
space dimensions of a D dimensional full object by reducing the space
dimension from non-zero to zero. Losing one space dimension, the D
dimensional full object becomes the ${\rm D - 1}$ dimensional core object. The
detached one-dimensional space becomes the radial space transverse to the ${\rm D - 1}$
 dimensional core object. The ${\rm D - 1}$ dimensional core object with the
transverse radial space constitutes the ${\rm D - 1}$ dimensional conformal field
theory (CFT).
\begin{equation}
\label{eq1}
\begin{array}{l}
 D\,\,{\rm dim}\ \ full\,\,object\,\, + \,{\rm D\,dim}\,\,empty\,\,\,object\,\,\left(
{\,AdS_{\rm D}}  \right)\,\,\buildrel {the\,\,\,decomposition\,} \over
\longrightarrow \,\, \\
{\rm  D\, -
\,1\,\,dim}\,\,\,core\,\,\,object\,\,with\,\,{\rm 1\,\,dim}\,\,transverse\,\,radial\,\,\,space\,
\\
 \, = \,\,{\rm D\, - \,1\,\,dim}\,\,\,conformal\,\,field\,\,\,theory\,\,\left(
{CFT} \right) \\
 \\
 \end{array}
\end{equation}

Therefore, this decomposition has the AdS/CFT duality (Maldacena, 1998;
Gubser, Klebanov and Polyakov, 1998; Witten, 1998), though this occurs in
tessellattice as AdS. In the decomposition theory for the AdS/CFT duality, $\rm  D
- 1$ dimensional CFT is the decomposition product from a D dimensional full
object in tessellattice as AdS$_{\rm D}$.

Furthermore, tessellattice decomposes not only the D dimensional background
space-time but also internal dimensional space-time. The internal dimension
numbers for brane, string, particle, and empty object are 3, 2, 1, and 0,
respectively. In the D dimensional background space-time, they are denoted
as 3$_{\rm D}$, 2$_{\rm D}$, 1$_{\rm D}$, and 0$_{\rm D}$.
The decompositions for them are as follows
\begin{equation}
\label{eq2}
\begin{array}{l}
 \,\,3_{\rm D} \,\, + \,\,0_{\rm D} \,\buildrel {the\,\,\,decomposition\,} \over
\longrightarrow \,\,2_{\rm D\, - \,1} \,1_{1} \, \\
 \,\,2_{\rm D} \,\, + \,\,0_{\rm D} \,\buildrel {the\,\,\,decomposition\,} \over
\longrightarrow \,\,1_{\rm D\, - \,1} \,1_{1} \, \\
 \end{array}
\end{equation}

The object in 1$_{1}$ (the transverse radial space) is a virtual particle
corresponding to the inerton surrounding the core object. A D-dimensional
full object in AdS is decomposed into a ${\rm D - 1}$ dimensional core object
surrounded by a virtual particle.

One example of the degenerate state of tessellattice is the universe before
the big bang or a series of small big bangs, which will be treated in detail
in the other work. As shown later, the most primitive part of the multiverse
is the primitiverse (Chung, 2002a). The primitiverse consists of closely
pack 10D strings, denoted as 2$_{10}$. There is no gap among 10D strings in
the primitiverse. It is the same everywhere. The primitiverse is flat with
$\Lambda  = 0$.

The combination of \textit{n} units of strings forms a loop, which then
obtains additional space dimension to form 11D brane, assigned as 3$_{11}$.
The additional space dimension for both background space-time and internal
space-time is in the form of de Sitter (dS) space. dS space has a function
of space expansion ($\Lambda  > 0$) that can expand one of the space
dimension in both the background space-time and the internal space-time from
zero to non-zero. Empty space as AdS space with a function of negative
expansion ($\Lambda  < 0$) is generated to balance dS, resulting in $\Lambda
 = 0$
\begin{equation}
\label{eq3}
\begin{array}{l}
 \left( {2{\kern 1pt} _{10}}  \right)_{{\kern 1pt} n} \,\,\, +
\,\,\,dS{\kern 1pt} _{10} \,\,\buildrel {the\,\,composition} \over
\longrightarrow \,\,3_{{\kern 1pt} 11} + \,\,AdS{\kern 1pt} _{11} \,\, =
\,\,\,\left( {\,0_{{\kern 1pt} 11}}  \right){\kern 1pt} _{n} \, \\
\,\,\,\,\,\,\,\,\,\,\,\,\,\,\,\,\,\,\,\,\,\,\,\,\,\,\,\,\,\,\,\,\,\,
\,\,\,\,\,\,\,\,\,\,\,\,\,\,\,\,\,\,\,\,\,\,\,\,\,\,\,\,\,\,\,\,\,\,
\,\,\,\,\,\,\,\,\,\,\,\,\,\,\,\,\,\,\,\,\,\,\,\,\,\,tessellattice
\\
 \end{array}
\end{equation}

The duality in Eq. (\ref{eq3}) is the dS/bulk duality. In the composition theory for
the dS/bulk duality, a D + 1 brane as bulk in AdS$_{\rm D+1}$ is the composition
product from D dimensional strings in dS$_{\rm D}$.

Normally, the decomposition of 3$_{11}$ results in \textit{n} units of
2$_{10}$ to complete the reversible process of the composition and the
decomposition. In some abnormal cases, the decomposition leads to the
AdS/CFT duality
\begin{equation}
\label{eq4}
\,\,3_{{\kern 1pt} 11} \,\,\,\,\,\, + \,\,\,\,\,\,\,\left( {\,\,0_{{\kern
1pt} 11}}  \right)_{{\kern 1pt} n} \,\,\,\,\,\,\,\,\,\,\,\buildrel
{the\,\,decomposition\,} \over \longrightarrow \,\left(
2{\kern 1pt} _{10}  \right)_ {n} \,\,\left(1{\kern 1pt} _{1}
 \right){\kern 1pt} _{n} .
\end{equation}

The full object, 3$_{11}$ in AdS decomposes into the core object, (2$_{10}$)
$_{n}$, surrounded by the virtual object, $\left( {1_{1}}  \right)_{n} .$
The virtual object is ``pregravity'' before a big bang. Pregravity is the
combination of the dS/bulk duality for the composition in Eq. (\ref{eq3}) and the
AdS/CFT in Eq. (\ref{eq4}) for the decomposition.

When the critical mass for the AdS/CFT duality reaches, the AdS/CFT duality
propagates in the primitiverse, and a universe is generated out of the
primitiverse as shown later.

As shown later, in the present light universe after the big bang, the object
structure is 4D particle, denoted as 1$_{4}$. \   (2$_{10}$)$_{n} $(1$_{1}$)$_{n
}$ in Eq. (\ref{eq4}) becomes (1$_{4}$)$_{n} $(1$_{1}$)$_{n}$  in the present
(observable) light universe
\begin{equation}
\label{eq5}
 (2{\kern 1pt} _{10})_n  (1{\kern
1pt} _{1})_n  \,\,\,\,\buildrel {\,bang} \over
\longrightarrow \,(1{\kern 1pt}_{4})_ n ( 1_ 1)_n  \,\, = \,\left( {\,1{\kern 1pt} _{4}
} \right)_{n} \,g
\end{equation}

The transverse radial space from 1$_{1}$ (virtual particle) is added to the
flat space in 1$_{4}$, resulting in the curved space for gravity in general
relativity. Thus, gravity as 1$_{1}$ is the origin of the Riemann tensor to
indicate the curved space wrapping around the flat core object.

\subsection*{3. The digital cosmic code for the space structure}

The cosmic digital code for the space structure consists of 1 and 0 for an
attachment space and a detachment space, respectively. Attachment space
attaches to object permanently with zero speed or reversibly at the speed of
light. Detachment space irreversibly detaches from the object at the speed
of light. Special relativity deals with the propagation speeds,
particularly, between zero and the speed of light.

As shown later, the universe starts with only attachment space without
detachment space (Chung, 2002a). The cosmic origin of detachment space is
the cosmic radiation that initiates big bangs. Some objects in 4D-attachment
space, denoted as 1$_{4}$, convert into the cosmic radiation in
4D-detachment space, denoted as 0$_{4}$
\begin{equation}
\label{eq6}
some\,\,objects\,\,\,in\,\,\,\,1{\kern 1pt} _{4} \quad \buildrel {\,bang}
\over \longrightarrow \,\,\,\,the\,\,cosmic\,\,radiation\,\,in\,\,\,0{\kern
1pt} _{4} \,
\end{equation}

Being massless particle, the cosmic radiation is on detachment space
continuously, and detaches from its own space continuously. The emergence of
the cosmic radiation allows the combination of \textit{n} units of
attachment space and \textit{n} units of detachment space. The combination
results in miscible space, binary partition space or binary lattice space.
\begin{equation}
\label{eq7}
\begin{array}{l}
 \left( {\,1_{4}}  \right)_{n} \,\,attachment\,space\,\,\,\,\, +
\,\,\,\,\left( {0_{4}}  \right)_{n} \,detachment\,\,space\,\,\,\,\buildrel
{combination} \over \longrightarrow \,\,\,\,\,\,\,\,\,\,\, \\  \\
 miscible\,space\,,\,\,\,\left( {1_{4} \,\,0_{4}}  \right)_{n\,}
binary\,lattice\,\,space, \ or,\ \\
 \qquad \qquad \qquad  \ \ \ \,\left( {1_{4}}  \right)_{n} \left(
{0_{4}}  \right)_{n} \,binary\,partition\,space \\
 \end{array}
\end{equation}

In miscible space, attachment space is miscible to detachment space, and
there is no separation of attachment space and detachment space. Binary
lattice space consists of repetitive units of alternative attachment space
and detachment space. Thus, binary lattice space consists of multiple
quantized units of attachment space separated from one another by detachment
space. Binary partition space consists of separated continuous phases of
attachment space and detachment space.

In miscible space, attachment space contributes zero speed, while detachment
space contributes the speed of light. A massless particle is on detachment
space continuously, and detaches from its own space continuously. For a
moving massive particle consisting of a rest massive part and a massless
part, the massive part with rest mass, $m_{0}$, is in attachment
space, and the massless part with kinetic energy, \textit{K}, is in
detachment space. The combination of the massive part in attachment space
and massless part in detachment leads to the propagation speed in between
zero and the speed of light.

To maintain the speed of light constant for a moving particle, the time
(\textit{t}) in moving particle has to be dilated, and the length
(\textit{L}) has to be contracted relative to the rest frame
\begin{equation}
\label{eq8}
\begin{array}{l}
 t\, = \,\, = \,{{t_{0}}  \mathord{\left/ {\vphantom {{t_{0}}  {\sqrt
{{\kern 1pt} {\kern 1pt} 1\, - \,\upsilon ^{2}\,/\,c^{2}}} }} \right.
\kern-\nulldelimiterspace} {\sqrt {{\kern 1pt} {\kern 1pt} 1\, - \,\upsilon
^{2}\,/\,c^{2}}} }\;\; = \;\;t_{0} \gamma , \\
 L\, = \,\,L_{0\,} /\,\gamma , \\
 E = K\, + \,m_{0} c^{2} = \,\gamma {\kern 1pt} m_{{\kern 1pt} 0}
\,c^{2}\,\,\,\, \\
 \end{array}
\end{equation}

\noindent
where $\gamma = 1/\sqrt {{\kern 1pt} 1 - \upsilon ^{2}/c^{2}} $ is the
Lorentz factor for time dilation and length contraction, \textit{E} is the
total energy and \textit{K} is the kinetic energy.

Binary lattice space consists of multiple quantized units of attachment
space separated from one another by detachment space. Binary lattice space
slices an object into multiple quantum states separated from one another by
detachment space. Binary lattice space corresponds to nilpotent universal
computational rewrite system (NUCRS) by Diaz and Rowlands (2003). NUCRS
starts with ``nothing'', and add new symbols, which must results in ``a zero
sum'' again. The addition of new symbols involves the sequential iterative
path from nothing (nilpotent) through conjugation, complexification, and
dimensionalization. Nilpotent corresponds to detachment space detached from
object.

In this paper, the universe starts with attachment space filled with objects
(information). The addition of detachment space results in binary lattice
space consisting of attachment space and detachment space. Detachment space
detaches from object. Detachment space contains no object that carries
information. Without information, detachment space is outside of the realm
of causality. Without causality, distance (space) and time do not matter to
detachment space. Thus, detachment space contains nothing (zero
information), so detachment space is non-local in terms of space-time.
Detachment can have any space and time. The non-locality of detachment space
in binary lattice space leads to the complete non-locality of binary lattice
space. Because detachment space contains no information, the non-locality of
binary lattice space cannot result in net new information. The changes of
objects in binary lattice space can only be expressed through conjugation,
complexification, and dimensionalization, which must result in zero sums,
indicating nothing (zero information) in detachment space (nilpotent) as in
NUCRS.

Basically, detachment space (nilpotent) de-localizes space-time in binary
lattice space. The non-local property of detachment space provides the
violation of Bell inequalities (Bell, 1964) in quantum mechanics in terms of
faster-than-light influence and indefinite property before measurement. The
non-locality in Bell inequalities does not result in net new information.

In binary lattice space, for every detachment space, there is its
corresponding adjacent attachment space. Thus, no part of the object can be
irreversibly separated from binary lattice space, and no part of a different
object can be incorporated in binary lattice space. Binary lattice space
represents coherence as wavefunction. Binary lattice space is for coherent
system.

Any destruction of the coherence by the addition of a different object to
the object causes the collapse of binary lattice space into binary partition
space.
\begin{equation}
\label{eq9}
\begin{array}{l}
 \left( {\left( {{\kern 1pt} {\kern 1pt} 0{\kern 1pt} _{4}}  \right)\;\left(
{{\kern 1pt} \;1{\kern 1pt} _{4}}  \right)} \right){\kern 1pt} _{n}
\qquad {\buildrel {collapse} \over \longrightarrow}  \qquad \left( {{\kern 1pt} {\kern
1pt} \;0{\kern 1pt} _{4}}  \right){\kern 1pt} _{n} \;\left( {{\kern 1pt}
{\kern 1pt} \;1_{{\kern 1pt} 4}}  \right){\kern 1pt} _{n} {\,\,\,}
\buildrel {mixing} \over \longrightarrow \, miscible\,space \\
binary\,\,lattice\,\,space\,\,\,\,\,\,\,\,\,\,\,\ \ \ \ binary\,\,partition\,\,space
\\
 \end{array}
\end{equation}

After the collapse, in binary partition space, attachment space attaches to
an object at any one location according to its probability without absolute
certainty, while detachment space separately detaches from all probability
density. By introducing a different object into an observed object,
experimental observation causes the collapse of binary lattice space. All
observations bring about binary partition space. Binary partition space is
the space immediately after the collapse of binary lattice space, and binary
partition space then changes to miscible space by mixing attachment space
and detachment space.

Binary lattice space can slice small object into quantum states, but binary
lattice space cannot slice a large object into quantum states due to
gravity. The difference between a small and a large object might be
associated with the comparison between the object's de Broglie wavelength
and the object size. Penrose (2000) pointed out that the gravity of large
object pulls different quantum states into one location. On the other hand,
the gravity of a small object is not strong enough to pull different states
into one location. Therefore, a large object is always in miscible space,
while a small object without outside interference is always in binary
lattice space.

The conventional explanation of the hidden extra space dimensions is the
compactification of the extra space dimensions. For example, six space
dimensions become hidden by the compactification, so space-time appears to
be four dimensional.

Bounias and Krasnoholovets (2003c) propose another explanation of the
reduction of $\rm > 4 D$ space-time into 4D space-time by slicing $\rm  > 4D$ space-time
into infinitely many 4D quantized units surrounding the 4D core particle.
Such slicing of $\rm > 4D$ space-time is like slicing 3-space D object into
2-space D object in the way stated by Michel Bounias as follows: ``You
cannot put a pot into a sheet without changing the shape of the ${\rm 2-D}$ sheet
into a ${\rm 3-D}$ dimensional packet. Only a ${\rm 2-D}$ slice of the pot could be a part
of sheet''.

This slicing of space-time dimension is done by detachment space indirectly.
As shown later, the direct slicing is the slicing of ``mass dimension''
derived from space-time dimension. The indirect slicing of $\rm  > 4D$ attachment
space by 4D detachment space is as follows
\begin{equation}
\label{eq10}
\begin{array}{l}
\,\,\,\,\,\,\left( {{\kern 1pt} 1_{4 + k}}
\right)_{m\,\,\,\,\,\,\,\qquad\qquad\quad  \buildrel {slicing} \over \longrightarrow }
\,\,\,\, \qquad  \left( {{\kern 1pt} \;1_{4}}  \right)_{m} \quad \,\,\,\,\quad\quad  + \quad
\,\,\sum\limits_{k = 1}^{k} {\left( {\left( {{\kern 1pt} {\kern 1pt} \;0_{4}
} \right)\;\left( {{\kern 1pt} {\kern 1pt} \;1_{4}}  \right)} \right)_{n,k}
} \\ >\,{\rm 4D}\,\,attachment\,\,space\,\,\,\,\,\,\,\,\,\,{\rm 4D}\,\,core\,\,attachment
\,\,space\,\,\,\,\,\,\,\,\,k\,{\kern
1pt} types\,\,of\,{\kern 1pt} {\kern 1pt} {\rm 4D} \,units\, \\
\,\,\,\,\,\,\,\,\,\,\,\,\,\,\,\,\,\,\,\,\,\,\,\,\,\,\,\,\,\,\,\,\,\,\,
\,\,\,\,\,\,\,\,\,\,\,\,\,\,\,\qquad\qquad\quad   k\,types\,of\,gauge\,force\,fields
\,\,in\,\,lattice\,space\,\,\,\,\,\,\,\,
\\
\end{array}
\end{equation}

The two products of the indirect slicing are the 4D-core attachment space
and 4D quantized units of attachment space separated by detachment space,
corresponding to binary lattice space. They are \textit{k} types of 4D
quantized units, representing the total number of space dimensions greater
than three-dimensional space. For example, the indirect slicing of 10D
attachment space produces 4D core attachment space and six types of 4D
quantized units. The value of \textit{n} approaches to infinite for
infinitely many 4D quantized units.

The core attachment space surrounded by infinitely many 4D quantized units
corresponds to the core particle surrounded by infinitely many small 4D
particles. The gauge force fields are made of such small 4D quantized
virtual particles surrounding the core particle.

Unlike tessellattice, binary lattice space for gauge force fields has no
boundary. It corresponds to the quantized asymptotically flat S-matrix with
$\Lambda $ = 0. Unlike tessellattice, binary lattice space for gauge force
fields is in low-density region. The gauge force fields (Chung, 2002a)
include electromagnetism, the strong force in the low-density pions, two
parity-nonconservation interactions, and two CP-nonconservation
interactions.

The transformation from 11D brane to 4D particle is AdS$_{5}$ X S$^{6}$
where AdS$_{5}$ is in tessellattice for gravity, S$^{6}$ is in binary
lattice space for gauge force fields, and there is no compactification of
extra space dimensions. In other words, one extra space dimension in the
seven extra space dimensions from 11D brane is reduced by tessellattice
(AdS) for gravity, and six extra space dimensions are reduced indirectly by
binary lattice space for the gauge force fields. With tessellattice and
binary lattice space, 11D brane is reduced to 4D particle surrounded by
gravity and the gauge force fields.

\subsection*{4. The cosmic dimension - varying dimension number}

The cosmic digital code is in the cosmic dimension, which is in the
framework of varying dimension number. Varying dimension number is derived
from varying speed of light (VSL) theory (Amelino-Camelia, 2001,2002;
Barrow, 2003; Ellis and Uzan, 2005; Magueijo, 2003). The constancy of the
speed of light is the pillar of special relativity. It is believed the
constancy of the speed of light takes place in the four dimensional
space-time whose space-time dimension number (four) is constant. In the
model of cosmology (Albrecht and Magueijo, 1999; Barrow, 1999, 2003) that
belongs to the VSL model, the speed of light varies in time. The time
dependent speed of light varies as some power of the expansion scale factor
\textit{a} in such way that
\begin{equation}
\label{eq11}
c\left( {t} \right)\, = \,c_{0} \,a^{n}
\end{equation}

\noindent
where $c_{0} > 0$ and \textit{n} are constants. The increase of speed of
light is continuous.

This paper posits quantized varying speed of light (QVSL), where the speed
of light is invariant in a constant space-time dimension number, and the
speed of light varies with varying space-time dimension number from 4 to 11.
In QVSL, the speed of light is quantized by varying space-time dimension
number
\begin{equation}
\label{eq12}
c_{\rm D} = c/\alpha ^{{\kern 1pt} \rm D\; - \;4},
\end{equation}

\noindent
where \textit{c} is the observed speed of light in the 4D space-time, $c_{\rm D}
$ is the quantized varying speed of light in space-time dimension number, D,
from 4 to 11, and $\alpha $ is the fine structure constant. Each dimensional
space-time has a specific speed of light. The speed of light increases with
the increasing space-time dimension number D. In the VDN model of cosmology,
the universe as the dual chiral universe that has the speed of light in 11D
space-time.

In special relativity, $E = M_{0} {\kern 1pt} c^{2}$ modified by Eq.
(\ref{eq12}) is expressed as
$$
E = M_{0} \cdot \left( {c^{2}/\alpha ^{{\kern 1pt} 2\left( {\rm D\; - \;4}
\right)}} \right)
\eqno(13a)
$$
$$
 = \left( {M_{0} /\alpha ^{{\kern 1pt} 2\;\left( {\rm d\; - \;4} \right)}}
\right)\; \cdot c^{{\kern 1pt} 2}.
\eqno(13b)
$$

Eq. (13$a$) means that a particle in the D dimensional space-time can have the
superluminal speed $c/\alpha ^{{\kern 1pt} \rm D - 4}$, which is
higher than the observed speed of light \textit{c}, and has the rest mass
$M_{0} $. Eq. (13$b$) means that the same particle in the 4D space-time with
the observed speed of light acquires $M_{0} /\alpha ^{{\kern 1pt} 2{\kern 1pt}
 \left( {\rm d - 4} \right)}$ as the rest mass, where $\rm d = D$. D
in Eq. (13$a$) is the space-time dimension number defining the varying speed
of light. In Eq. (13$b$), d from 4 to 11 is ``mass dimension number'' defining
varying mass. For example, for $\rm D = 11$ Eq. (13$a$) shows a superluminal
particle in eleven-dimensional space-time, while Eq. (13$b$) shows that the
speed of light of the same particle is the observed speed of light with the
4D space-time, and the mass dimension is eleven. In other words, 11D
space-time can transform into 4D space-time with 11d mass dimension. QVSL in
terms of varying space-time dimension number, D, brings about varying mass
in terms of varying mass dimension number, d.

The QVSL transformation transforms both space-time dimension number and mass
dimension number. In the QVSL transformation, the decrease in the speed of
light leads to the decrease in space-time dimension number and the increase
of mass in terms of increasing mass dimension number from 4 to 11,
$$
c_{{\kern 1pt}\rm D} = c_{{\kern 1pt}{\rm D}\; - \;n} /\alpha ^{{\kern 1pt} 2\;n},
\eqno(14a)
$$
$$
M_{0,\;{\rm D\;,\,d}} = M_{0,\;{\rm D}\; - \;n,\;\;{\rm d}\; + \;n} \alpha ^{{\kern 1pt}
2\;n},
\eqno(14b)
$$
$$
{\rm D,\,\,d} \ \  \buildrel {QVSL} \over \longrightarrow  \ \  \left( {{\rm D} \mp n}
\right),\,\,\,\left( {{\rm d} \pm n} \right)
\eqno(14c)
$$

\noindent
where D is the space-time dimension number from 4 to 11 and d is the mass
dimension number from 4 to 11. For example, the QVSL transformation steps a
particle with 11D4d to a particle with 4D11d. In terms of rest mass, 11D
space-time has 4d with the lowest rest mass, and 4D space-time has 11d with
the highest rest mass.

In such a way, the QVSL transformation is an alternative to the Higgs
mechanism to gain rest mass. In the QVSL, the speed of light is constant in
a specific space-time dimension number, such as 4 for our four-dimensional
space-time. In different space-time dimension numbers (from 4 to 11), speeds
of light are different. In our four-dimensional space-time, the speed of
light is the lowest, so according to special relativity ($E = M_{0} {\kern
1pt} c^{2}$), with constant energy, the rest mass in our four-dimensional
space-time is the highest. Thus, instead of absorbing the Higgs boson to
gain rest mass, a particle can gain rest mass by decreasing the speed of
light and space-time dimension number. The QVSL transformation also gains a
new quantum number, ``mass dimension number'' from 4 to 11 to explain the
hierarchical masses of elementary particles. Since the Higgs bosons have not
been found experimentally, the QVSL transformation to gain rest mass is a
good alternative. In terms of vacuum energy, the four-dimensional space-time
has the zero vacuum energy with the highest rest mass, while D$ > 4$ have
non-zero vacuum energy with lower rest mass than 4D. Since the speed of
light for a particle with dimension $> 4$D is greater than the speed of light
for a 4D particle, the observation of $> 4$D particles by 4D particles
violates casualty. Thus, particles with dimension $> 4$D should be treated as
hidden particles with respect to 4D particles. In general, particles with
different space-time dimensions are transparent and oblivious to one
another.

In the normal supersymmetry transformation, the repeated application of the
fermion-boson transformation carries over a boson (or fermion) from one
point to the same boson (or fermion) at another point at the same mass. In
the ``varying supersymmetry transformation'', the repeated application of
the fermion-boson transformation steps a boson from one point to the boson
at another point at different mass dimension number in the same space-time
number. The repeated varying supersymmetry transformation carries over a
boson B$_{\rm d}$ into a fermion F$_{\rm d}$ and a fermion F$_{\rm d}$ to a boson
B$_{\rm d-1}$, which can be expressed as follows
$$
M_{\rm d,\;F} \,\,\;\,\, = \,\,M_{\rm d,\;B} \;\alpha _{\rm d,\;B} ,
\eqno(15a)
$$
$$
M_{\rm d - \;1,\;B} = M_{\rm d,\;F} \;\alpha _{\rm d,\;F} ,
\eqno(16b)
$$
where \textit{M}$_{\rm d, B}$ and \textit{M}$_{\rm d, F}$ are the masses for a boson
and a fermion, respectively, d is the mass dimension number, and $\alpha
_{\rm d,\;B} $ or $\alpha _{\rm d,\;F} $ is the fine structure constant that is the
ratio between the masses of a boson and its fermionic partner. Assuming
$\alpha _{\rm d,\;B} $ or $\alpha _{\rm d,\;F} $, the relation between the bosons in
the adjacent dimensions then can be expressed as
$$
M_{\rm d - \;1,\;B} = M_{\rm d,\;B} \;\alpha _{\rm d}^{2} .
\eqno(15c)
$$

Eqs. (15) show that it is possible to describe mass dimensions $> 4$ in the
following way
$$
\rm  F_{5} \,B_{5} \,F_{6} \,B_{6} \,F_{7} \,B_{7} \,F_{8} \,B_{8} \,F_{9}
\,B_{9} \,F_{10} \,B_{10} \,F_{11} \,B_{11} ,
\eqno(16)
$$

\noindent
where the energy of B$_{11}$ is the Planck energy. Each mass dimension
between 4d and 11d consists of a boson and a fermion. Eqs. (15) show a
stepwise transformation that converts a particle with d mass dimension to d
$ \pm $ 1 mass dimension. The transformation from a higher dimensional
particle to the adjacent lower dimensional particle is the fractionalization
of the higher dimensional particle to the many lower dimensional particle in
such way that the number of lower dimensional particles becomes $n_{{\kern
1pt} \rm d - 1} = n_{{\kern 1pt}\rm  d} /\alpha ^{{\kern 1pt} 2}$. The
transformation from lower dimensional particles to higher dimensional
particle is a condensation. Both the fractionalization and the condensation
are stepwise. For example, a particle with 4D (space-time) 10d (mass
dimension) can transform stepwise into 4D9d particles. Since the
supersymmetry transformation involves translation, this stepwise varying
supersymmetry transformation leads to a translational fractionalization and
translational condensation, resulting in expansion and contraction. At the
same time it should be mentioned that research by Krasnoholovets (2000)
points out to the fact that only fermions are true canonical particles,
while bosons are rather combined particles consisting of fermions.

Another type of the varying supersymmetry transformation is not stepwise. It
is the leaping varying supersymmetry transformation that transforms a
particle with d mass dimension to any d $ \pm $ \textit{n} mass dimension.
The transformation involves the slicing-fusion of particle. The
transformation from d to $\rm d - $\textit{n} involves the slicing of a particle
with d mass dimension into two parts: the core particle with $\rm d - $\textit{n}
dimension and the \textit{n} dimensions that are separable from the core
particle. Such \textit{n} dimensions are denoted as \textit{n} ``dimensional
orbitals'', which become force fields (Chung 2002a). The sum of the number
of mass dimensions for a particle and the number of dimensional orbitals
(DO's) is equal to 11 for all particles with mass dimensions. Therefore,
$$
{\rm F_{d}} \, = \,{\rm F}_{{\rm d} - n} \, + \,\,\left( {11\, - \,
{\rm d}\, + \,n} \right)\,{\rm DO's}
\eqno(17)
$$

\noindent
where ${\rm 11 - d }+ n$ is the number of dimensional orbitals (DO's) for
${\rm F}_{{\rm d} - n}$. For example, the slicing of 4D9d particle produces 4D4d
particle that has d = 4 core particle surrounded by 7 separable dimensional
orbitals in the form of \\
\begin{center}
B$_5$F$_5$B$_6$F$_6$B$_7$F$_7$B$_8$F$_8$B$_9$F$_9$B$_10$F$_10$B$_11$.
\end{center}
Since the slicing process is not stepwise from higher mass dimension to
lower mass dimension, it is possible to have simultaneous slicing. For
example, 4D9d particles can simultaneously transform into 4D8d, 4D7d, 4D6d,
4D5d, and 4D4d core particles, which have 3, 4, 5, 6, and 7 separable
dimensional orbitals, respectively.

Therefore, varying supersymmetry transformation can be stepwise or leaping.
Stepwise supersymmetry transformation is translational fractionalization and
condensation, resulting in stepwise expansion and contraction. Leaping
supersymmetry transformation is not translational, and it is slicing and
fusion, resulting possibly in simultaneous formation of different particles
with separable dimensional orbitals.

In summary, the QVSL transformation carries over both space-time dimension
number and mass dimension number. The varying supersymmetry transforms
varying mass dimension number in the same space-time number as follows (once
again, D = space-time dimension number and d = mass dimension number).
\[
\begin{array}{l}
{\rm  D,\,\,d}  \qquad\qquad \qquad  \  \buildrel {QVSL} \over \longrightarrow
\qquad\qquad\qquad \ \ \  \left( {{\rm D} \mp n}
\right),\,\,\,\left( {{\rm d} \pm n} \right) \\
{\rm D,\,\,d} \quad \ \buildrel {stepwise\,\,\,\,\,varying\,\,\,\,supersymmetry\,\,} \over
\longrightarrow \quad \ \ {\rm D},\,\,\left( {{\rm d} \pm 1} \right)\, \\
{\rm  D,\,\,d} \qquad  \buildrel {leaping\,\,\,\,varying\,\,\,supersymmetry} \over
\longrightarrow  \qquad  {\rm D},\,\left( {{\rm d} \pm \,n} \right) \\
 \end{array}
\]

\subsection*{5. Cosmology}

The cosmic gene controls the maturation processes of different cosmic organs
as different universes. The universal maturation process of our dual Vedic
universe involves four stages: the primitiverse, the dual chiral universe,
the dual achiral universe, and the dual Vedic universe. Different stages of
the universal maturation process are different cosmic genetic expressions.
The emergencies of empty object, achirality, and zero vacuum energy lead to
the dual chiral universe, the dual achiral universe, and the dual Vedic
universe, respectively as follows

\[
\begin{array}{l}
\  \qquad  primitiverse\,\;\buildrel {empty\,\,\,object} \over \longrightarrow
\;\, \quad dual\,\,\,chiral\,\,universe\,\;   \\  \quad\qquad\qquad \qquad\qquad
\buildrel {achirality} \over \longrightarrow \,\;
\qquad  dual\,\,achiral\,\,universe  \\
\quad\qquad \quad\qquad \quad  \buildrel {zero\,\,vacuum\,\,energy} \over \longrightarrow
 \quad dual\,\,expanding\,\,universe \\
 \end{array}
\]

The primitiverse consists of the closely packed 10D string-antistrings. The
primitiverse with $\Lambda $ = 0 does not have empty object and detachment
space. The dual chiral universe started from the combination of \textit{n}
units of 10D strings into 11D brane as in Eq. (\ref{eq3}). Through symmetry, the
combination of \textit{n} units of antistrings forms 11D anti-brane. The
combination of 3$_{11}$ and 0$_{11}$ brings about 2$_{10}$ 2$_{1}$, where
2$_{1}$ is pregravity, \textit{g}. In the same way, the combination of
3-$_{11}$ and 0$_{11}$ brings about $2_{ - 10}  \quad 1_{ - 1} $, where $1_{ -
1} $ is anti-pregravity,\textit{} $g_{ -}  $.
$$
\begin{array}{l}
 \left( {2_{10}}  \right)_{n} \,\,\, + \,\,\,dS \ \ \,\qquad\qquad  \buildrel {composition\,}
\over \longrightarrow \, \ \quad \quad \,3_{11} \,\, + \,\,\left( {\,0_{11}}  \right)_{n}
\\
 \left( {2_{ - 10}}  \right)_{n} \,\,\, + \,\,\,dS\, \  \qquad\qquad  \buildrel
{composition} \over \longrightarrow \, \ \quad\quad \,3_{ - 11} \,\, + \,\,\left( {\,0_{
- 11}}  \right)_{n} \, \\
 3_{11} \,\,\, + \,\,\,\,\,\left( {0_{11}}  \right)_{n} \ \ \ \ \quad\qquad  \buildrel
{decomposition\,} \over \longrightarrow \, \quad\qquad  \left( 2_{11}  \right)_{n} \left( 1_{1}
\right)_{1} \, = \,\,\left( {\,2_{10}}  \right)_{n} \,g \\
 3_{ - 11} \,\,\, + \,\,\,\left( {0_{ - 11}}  \right)_{n} \, \qquad\quad \ \buildrel
{decomposition\,} \over \longrightarrow \, \quad\quad  \left(2_{ - 10}  \right)_{n}
\;\left(1_{ - 1} \right)_{n} \, = \,\,\left( {\,2_{ - 10}}  \right)_{n} \,g_{
-}  \\
 \left( {2_{10}}  \right)_{n} \,g\,\,\,\, + \,\,\,\left( {2_{ - 10}}
\right)_{n} \,g_{ -}  \qquad  \,\,\buildrel {} \over \longrightarrow \, \quad\quad \ \ \ \
\left({\,2_{10}}  \right)_{n} \,g\,g_{ -}  \,\,\left( {2_{ - 10}}  \right)_{n} \\
 \end{array}
\eqno(18)
$$

Pregravity is repulsive to anti-pregravity, so (2$_{10}$)$_{n}$ and
(2$_{-10}$)$_{n}$ are separated from each other. When the critical mass of
the AdS/CFT duality reaches, the AdS/CFT duality propagates in the
primitiverse continuously, resulting in a large number for n.

Other than pre-gravity, there are two other forces: the pre-strong force and
the pre-charged force, as the predecessor of the strong force and
electromagnetism, respectively. Both of them are from the quantized
vibration of the strings. The original force in the primitiverse is the
short-ranged pre-strong force, ``s'', among strings. It is from the
reversible process of the absorption and the emission of the massless
particles among the strings. The pre-strong force also forms the bond
between the primitiverse and the strings with gravity. The vibration of the
string with gravity generates the long-ranged pre-charged force, ``e'', as
the reversible process of the absorption and the emission of the massless
particles among the strings with gravity. The string with gravity has the
positive pre-charged force, and the antistring with antigravity has the
negative pre-charged force. The positive pre-charged force is attractive to
the negative pre-charged force.

The appearances of different forces follow the specific sequence. In the
sequence, the pre-strong force exists first. Then, the emergence of the
repulsive force between pregravity and anti-pregravity forces the string and
the antistring to move away from each other. All of the strings go to one
domain with pregravity, while all of the antistrings go to the opposite
domain with anti-pregravity. Subsequently, the pre-strong force connects the
newly formed strings or the antistring with the previously formed strings or
antistrings. Finally, the pre-charged force emerges. The domain occupied by
the strings is opposite from the domain occupied by the antistrings, so the
strings and the antistrings are chiral. This specific sequence provides the
formation of the dual chiral universe.
$$
{\rm primiti \ verse} \ \,s\,\,\left( {2_{10} \,s\,\,2_{10} \,s} \right)_{n}
{\kern 1pt} {\kern 1pt} \;\frac{{g^{ +} }}{{e^{ +} }}\,\;\frac{{g^{ -
}}}{{e^{ -} }}\,\,\left( {2_{ - 10} \,s \ \,2_{ - 10} \,s} \right)_{n}
s\,\,{\rm primiti \ verse}
\eqno(19)
$$

\noindent
where \textit{g} is the pregravity, \textit{e} is the pre-charged force,
\textit{s} the pre-strong force and \textit{n} numbers of repetitive units.

The dual chiral universe consists of two universes: 10D strings with
positive energy and gravity and 10D antistrings with negative energy and
anti-pregravity. The two universes are opposite (chiral) in CP, energy, and
pregravity.

During the steady conversion from the primitiverse to the dual chiral
universe takes place, the total volume of the primitiverse and the dual
chiral universe remain constant. To maintain this constant volume, the
attractive force (\textit{A}) between the positive and negative precharged
forces is equal to the sum of the repulsive force (\textit{R}) between
pregravity and anti-pregravity, and the special global short-ranged
pre-strong force (\textit{C}) connecting the dual chiral universe and the
primitiverse. ${A = R + C}$ is a non-localized global relation for the
constant total volume of the universes. If ${A > R + C}$, the total
volume is smaller, and if ${A < R + C}$, the total volume is larger.

There is a small amount of the abnormal sequence of the appearance of force
in the dual chiral universe. In the abnormal development sequence, the
pre-strong force exists first. Then, the emergence of pregravity occurs
simultaneously with the attractive force from the pre-charged forces,
drawing the string and the antistring together,
$$
g^{ +} \left( {2_{10}}  \right)_{m} \,e^{ +} \,e^{ -} \left( {\,2_{ - 10}}
\right)_{m} g^{ -}
\eqno(20)
$$

The combined string-antistring units go impartially to either side of the
dual chiral universe, resulting in the achiral string-antistring units.
(Essentially, attractive force and repulsive force are the tools to form
chirality and achirality.) In the universe, local interactions are either
chirality-specific or achirality-specific. Unable to interact with the
region inside the dual chiral universe, the achiral string-antistring units
are separated from the dual chiral universe, and congregate in the area
connecting the primitiverse and the dual chiral universe. They form the
achiral domain next to the primitiverse.
$$
{\rm primiti \,verse} \ \,s\,  \ \,g + \left( {2_{10}}  \right)_{m} \;e^{ +} \;e^{
-} \left( {\,2_{ - 10}}  \right)_{m} \;g^{ -} \ \ \,s\, \ {\rm primiti \, verse}
\eqno(21)
$$

\noindent
where \textit{m} is much smaller than \textit{n} from Eq. (19).

Such achiral domain connects with the primitiverse, but does not connect
with the dual chiral universe. The result is the decrease of the connection
between the dual chiral universe and the primitiverse. However, as a
non-localized global relation, $A = R + C$ continues with the right
amount of \textit{C} contributed by the primitiverse as long as there is
still connection between the primitiverse and the dual chiral universe.

As the dual chiral universe grows, the achiral domain also grows.
Eventually, the dual chiral universe is disconnected completely from the
primitiverse by the achiral domain. Without \textit{C}, the excess
attractive force ($A > R$) between positive charged strings and
negative charged antistrings causes the dual chiral universe to collapse,
and the repulsive force between pregravity and anti-pregravity causes the
dual chiral universe to inverse. As the 10D-strings and the 10D-antistrings
move toward each other, the 10D-strings and the 10D-antistrings turn inside,
and pregravity and anti-pregravity turn outside. The ``gulf'' separates the
dual chiral universe and the primitiverse forms. Eventually, the 10D-strings
and the 10D-antistrings coalesce.

During the coalescence, the two chiral universes coexist in the same
space-time, which is predicted by the Santilli isodual theory (Santilli,
2005). Anitparticle for our positive energy universe is described by
Santilli as follows, ``this identity is at the foundation of the perception
that antiparticles ``appear'' to exist in our space, while in reality they
belong to a structurally different space coexisting within our own, thus
setting the foundations of a ``multidimensional universe'' coexisting in the
same space of our sensory perception'' (Santilli, 2005, p. 94).
Antiparticles in the positive energy universe actually come from the
coexisting negative energy universe. With chiral symmetry, the isodual
theory describes the coexistence of the two chiral universes.

The coexisting chiral universes do not remain chiral. The mixing of the two
chiral universes results in the two achiral universes. The mixing process
follows the isodual hole theory that is the combination of the isodual
theory and the Dirac hole theory. In the Dirac hole theory that is not
symmetrical, the positive energy universe has an unobservable infinitive sea
of negative energy. A hole in the unobservable infinitive sea of negative
energy is the observable positive energy antiparticle. The isodual hole
theory has two symmetrical sets of coexisting holes in the symmetrical
coexisting seas of positive energy and negative energy.

In the dual chiral universe, one universe has positive energy, strings, and
pregravity, and one universe has negative energy, antistrings, and
anti-pregravity. During the mixing when two chiral universes coexist, a half
of antistrings in the negative energy universe moves to the positive energy
universe, and the process leaves the Dirac holes in the negative energy
universe. The antistrings moved become positive energy antistrings in the
positive energy universe. In the same way, a half of strings in the positive
energy universe moves to the negative energy universe, and the process
leaves the Dirac holes in the positive energy universe. The strings moved
become negative energy strings in the negative energy universe. The result
is that a half of antistrings moves from the negative energy universe to the
positive energy universe to become positive energy antistrings, and a half
of strings from the positive energy universe to the negative energy universe
to become negative energy strings.

Both positive energy universe and negative energy universe have
strings-anti-strings. The universes become achiral in terms of strings and
antistrings. With the odd number (\ref{eq9}) of space dimension, 10D string has
chiral symmetry. Chiral 10D string cannot survive in the achiral universes,
so 10D strings-antistrings become 10D particles-antiparticles. 10D
particles-antiparticles have the multiple dimensional Kaluza-Klein structure
with flexible requirement for space dimension number, so they do not need to
have a fixed space dimension number. Thus, the isodual hole theory converts
the chiral universes into the achiral universes, and strings-antistrings
with fixed space dimension number into particles-antiparticles with varying
space dimension number.

The dual achiral universe consists of positive energy
particles-antiparticles with pregravity and negative energy
particles-antiparticles with anti-pregravity. The two coexisting universes
are represented as below:
$$
\begin{array}{l}
{\rm primiti  \, verse} \quad \left[ {\rm gulf} \right]\,\;1/2g^{ +} \left( 1_{10}
{\kern 1pt} 1_{ - 10}  \right)_{n/2} \,\,1/2g^{ +} \ \left[ {\rm {gulf}}
\right]\quad {\rm primiti \, verse} \\
\,\,\,\,\,\,\,\,\,\,\,\,\,\,\,\,\,\,\,\,\,\,\,\,\,\,\,\,
\,\,\,\,\,\,\,\,\,\,\,\,\,\,\,\,\,\,\,\,\,{ coexisting \,\, with}
\\
{\rm  primiti \, verse}\quad \left[ {\rm gulf} \right] \ 1/2g^{ -} \left( 1_{10}
{\kern 1pt} 1_{ - 10}  \right)_{n/2}
 \ 1/2g^{-} \ \left[ {\rm gulf} \right]  \  {\rm  primiti \, verse}.
\end{array}
\eqno(22)
$$

The dual achiral universe consists of four equal parts: two groups of
achiral particle-antiparticles, pregravity, and anti-pregravity

The size of the dual chiral universe is determined by the ratio between the
number of the chiral units and the number of the achiral units. The
primitiverse and the dual achiral universe are different in the composition
of objects and spaces, and are separated from each other permanently.
Consequently, the two universes are completely transparent and oblivious to
each other.

Without relation with the primitiverse, the dual achiral universe has its
own vacuum energy that decreases from the non-zero in the primitiverse to
zero. With decreasing vacuum energy and the Kaluza-Klein structure without a
fixed number of space dimensions, the space-time dimension and the mass
dimension of particle-antiparticles decrease to lower dimensional space-time
and lower dimensional mass. The decrease to lower mass dimension results in
the fractionalization of particle-antiparticles into lower mass particles,
leading to the expansion of the universe. The result is the dual Vedic
universe in the two different modes: the slow mode for the dark universe and
the quick mode for the light universe. As the space energy decreases, the
mass-energy is created, corresponding to the cosmology in the Vedas (Roy,
1999). In the cosmology described by the Vedas, the universe starts with the
void, i.e. the web of space does not have energy, or force. The creation of
mass-energy occurs on the surface of the void web, resulting in the
expansion of space, or the universe. In dual Vedic universe, the universe
starts with the high space energy, leading to the nearly void with nearly
zero mass-energy. Thus, the dual universe is called the dual Vedic universe.
As shown later, the dark universe without detachment space has no light,
while the light universe with detachment space has light.

In the slow mode, the vacuum energy decreases to zero gradually, and the
space-time dimension of the 10D particle-antiparticle with antigravity
decreases from 10D to 4D, stepwise. In the dark universe, the 10D4d
particles at high vacuum energy transform into 9D5d particles at low vacuum
energy through the QVSL transformation. Through the varying supersymmetry
transformation, 9D5d becomes 9D4d. Such varying supersymmetry transformation
brings about the stepwise translational fractionalization, resulting in
cosmic expansion. Further decrease in vacuum energy repeats the same process
again until particles are the 4D particles at zero vacuum energy as follows

\[
\begin{array}{l}

{\textsl{The\,Slow\,Mode:\,\,
 The\,Hidden\,Dark\,\,Universe\,\,and\,\,the\,Dark\,\,Energy\,\,Universe}}
\\
 \\
\qquad{\rm  10D4d} \to {\rm 9D5d} \to {\rm 9D4d} \to {\rm 8D5d} \to
{\rm 8D4d} \to {\rm 7D5d} \to  \bullet \bullet
\bullet \bullet   \\
\qquad \to {\rm 5D4d} \to \, {\rm 4D5d}  \, \to {\rm 4D4d}  \  \mapsto
\,the \ hidden \ dark \ universe   \\
\qquad  \leftarrow \mapsto \,\,dark \  energy\, \leftarrow \\
 \end{array}
\]

The dark universe consists of two periods: the hidden dark universe and the
dark energy universe. The hidden dark universe composes of the $> 4$D
particles. As mentioned before, since the speed of light for $> 4$D particle
is greater than the speed of light for 4D particle, the observation of $> 4$D
particles by 4D particles violates casualty. Thus, $> 4$D particles are hidden
particles with respect to 4D particles. The universe with $> 4$D particles is
the hidden dark universe. The 4D particles transformed from hidden $> 4$D
particles in the dark universe are observable dark energy for the light
universe, resulting in the accelerated expanding universe. The accelerated
expanding universe consists of the positive energy 4D
particles-antiparticles and dark energy that includes the negative energy 4D
particles-antiparticles and the antigravity. Dark energy does not contradict
to Santilli's (2005) isodual theory (the combination of the isodual theory
and the hole theory), where two universes coexist. Since the dark universe
does not have detachment space, the presence of dark energy is not different
from the presence of the high vacuum energy.

The quick mode is used in the light universe. Through zero vacuum energy,
10D4d particle transforms through the quick QVSL transformation quickly into
4D10d particles. 4D10d particle then transforms and fractionalizes quickly
through varying supersymmetry transformation into 4D9d, resulting in an
inflationary expansion (Guth 1981; Linde, 1982; Albrecht and Steinhardt,
1982; Chung, 2002a). The inflationary expansion occurs between the energy
for 4D10d =$E_{{\kern 1pt}\rm Planck} {\kern 1pt} \alpha ^{{\kern 1pt} 2} = 6
\times 10^{14}$ GeV and the energy for 4D9d = E$_{10}{\kern 1pt} \alpha ^{{\kern 1pt}
2} = 3 \times 10^{{\kern 1pt} 10}$ GeV.

At the end of the inflationary expansion, all 4D9d particles undergo
simultaneous slicing to generate equally by mass and number into 4D9d, 4D8d,
4D7d, 4D6d, 4D5d, and 4D4d core particles. Baryonic matter is 4D4d, while
dark matter consists of the other five types of particles (4D9d, 4D8d, 4D7d,
4D6d, and 4D5d). The mass ratio of dark matter to baryonic matter is 5 to 1
in agreement with the observation (Rees, 2003) showing the universe consists
of 25\% dark matter, 5\% baryonic matter, and 70\% dark energy. Dark matter
contributes to the inhomogeneous structure of baryonic matter (Chung 2002a).

The mechanism for the simultaneous slicing of mass dimensions requires
detachment space that slices mass dimensions. The dual achiral universe
consists of 10D particle-antiparticle. With the CP symmetry, 10D
particle-antiparticle undergoes annihilation (implosion). Annihilation is
the detachment of energy from the original position. The space is detachment
space, and the detached energy is cosmic radiation. The particles with CP
asymmetry remain as the particles (matter). The whole process becomes

\bigskip

\[
\begin{array}{l}
 \textsl{The \ Quick \,Mode: \,The \, Light \, Universe} \\
 \\
{\rm 10D4d} \quad \buildrel {quick\,\,QVSL\,\,transformation} \over \longrightarrow
\quad  {\rm 4D10d} \quad  \buildrel {stepwise\,varying\,supersymmetry,\,\,inflation} \over
\longrightarrow \\
{\rm 4D9d} \quad  \buildrel {simultaneous\,\,\,slicing} \over \longrightarrow
\ \ dark\,matter\,\left( {\rm 4D9d + 4D8d + \;4D7d + 4D6d + 4D5d} \right) \\
+ \,baryonic\,\,matter\,\left( {\rm 4D4d} \right) +
\,cosmic\,\,radiation     \\
\to \;thermal\,cosmic\,expansion\,\left(
{the\,big\,bang} \right) \\
 \end{array}
\]

\bigskip

For baryonic matter, the slicing of mass dimensions is as follows
$$
\begin{array}{l}
 \,\,\,\,\,\,\,\,\,\,\,\,\,\,\left( {1_{4 + 6}}  \right)_{m}
{\qquad\qquad \ \ \  \buildrel {slicing} \over \longrightarrow}
\,\,\,\,\,\,\,\quad   \left( {\kern 1pt}
1_{4}  \right)_{m} \quad\qquad\qquad  + \quad\qquad  \sum\limits_{1}^{6} {\left( {\left(
{{\kern 1pt} 0_{4}}  \right)\;\left( {{\kern 1pt} 1_{4}}  \right)}
\right)_{n,6}}  \\
 {\rm 4D >
\,4d} \,attach.\,\,space\,\,\,\,\,\,\,\,\,\,\,
{\rm 4D4d}\,core\,\,attach.\,space\,\,\,\,\,\,6\,\,types\,{\rm 4D4d}\,\,units\,
\\
 \end{array}
\eqno(23)
$$

\noindent
where 4 and 6 (for six gauge force fields) are d mass dimensions.

\begin{figure}
\begin{center}
\includegraphics[scale=0.7]{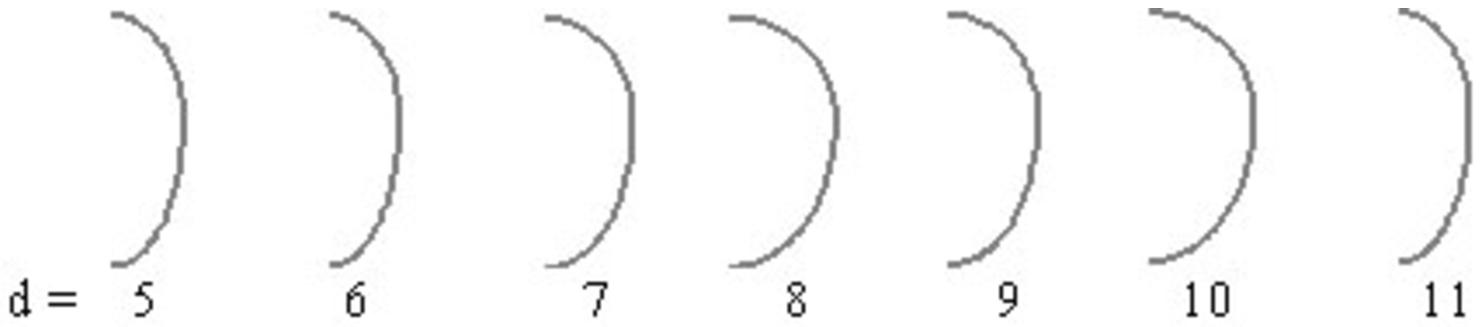}
\caption{\small{The force fields as $> 4$d mass dimensions
(dimensional orbitals)}}
\label{Figure 1}
\end{center}
\begin{center}
\includegraphics[scale=0.6]{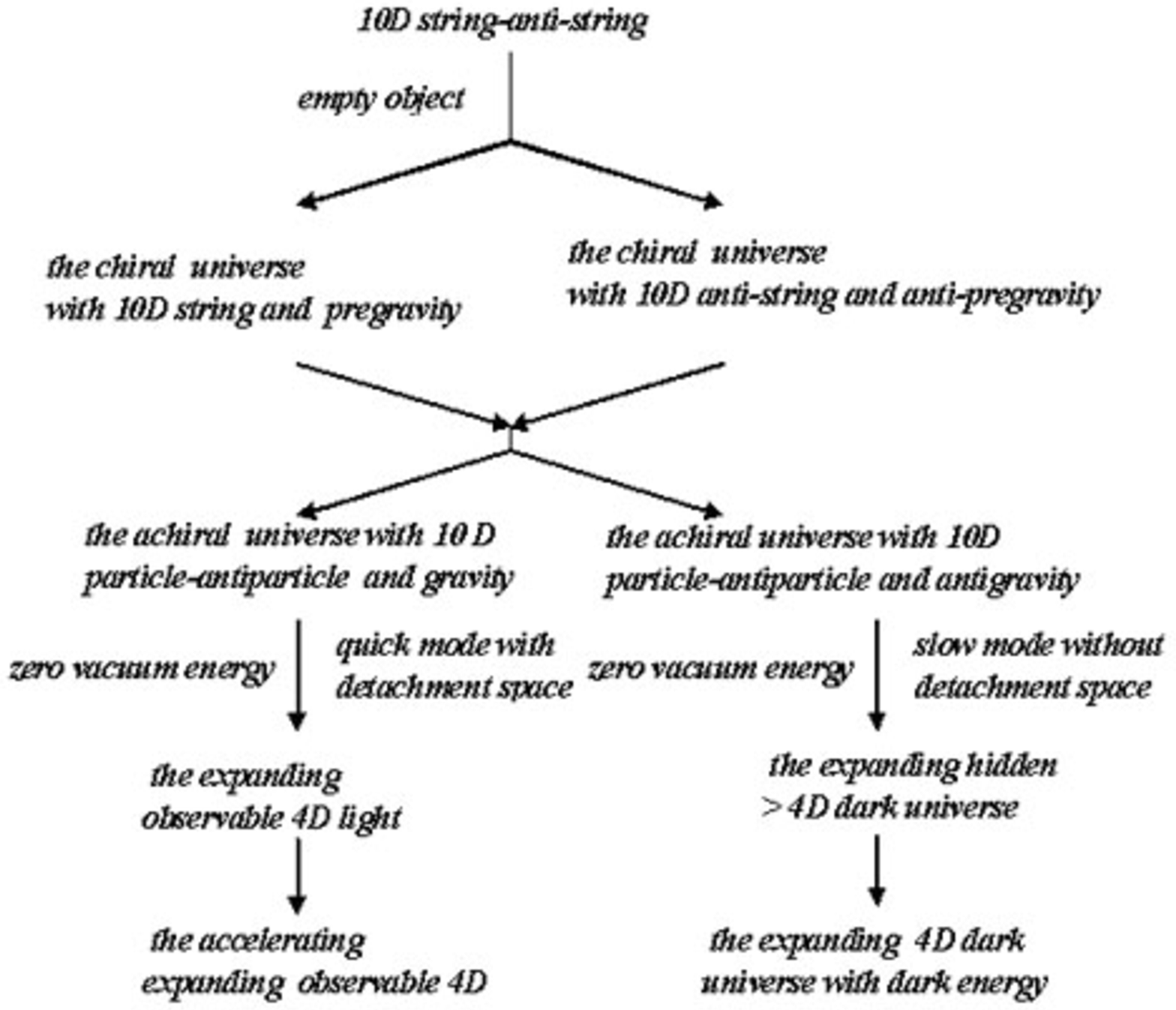}
\caption{\small{Cosmology.}}
\label{Figure 2}
\end{center}
\end{figure}

The two products of the slicing are the 4D4d-core attachment space and six
types of 4D4d quantized units. The 4D4d core attachment space surrounded by
six types of infinitely many 4D4d quantized units corresponds to the core
particle surrounded by six types of infinitely many small 4D4d particles.
The gauge force fields are made of such small 4D4d quantized virtual
particles surrounding the core particle.

The six $> 4$d mass dimensions (dimensional orbitals) for the gauge force
fields and the one mass dimension for gravity are as in Figure 1.

The dimensional orbitals form the base for the periodic table of elementary
particles to calculate the masses for graviton, gauge bosons, quarks, and
leptons (Chung, 2002a).

The summary of cosmology is shown in Figure 2.

\bigskip

\subsubsection*{6. Force Fields}

The two types of force fields are the force fields in tessellattice and the
force fields in binary lattice space. Tessellattice consisting of empty
objects corresponds to AdS ($\Lambda  < 0$) in high-density region with
boundary. The force fields derived from the AdS/CFT duality include the
original gravity before the big bang and the strong force in nucleus.

Binary lattice space, consisting of alternative attachment space and
detachment space corresponds to S-matrix ($\Lambda  = 0$) in low-density
region without boundary. The force fields derived from the 4D quantized
units of their respective mass dimensions include electromagnetism, the
strong force in pion, and the weak interaction (Chung, 2002a).

The table below lists the two types of force fields.

\textbf{Table.} Force fields.

{\small{
\newcommand{\PreserveBackslash}[1]{\let\temp=\\#1\let\\=\temp}
\let\PBS=\PreserveBackslash
\begin{longtable}
{|p{174pt}|p{190pt}|}
a & a  \kill
\hline
\underline {high density region with boundary}&
\underline {tessellattice (AdS) $\Lambda $ < 0} \\
\hline
\underline {}original gravity before the big bang&
The dS/bulk duality and the AdS/CFT duality \\
\hline
the strong force in baryon&
the CFT/AdS duality, the AdS/CFT duality, and the AdS/SU(\ref{eq3}) duality \\
\hline
&
 \\
\hline
\underline {low density region without boundary}&
\underline {binary lattice space (S-matrix) $\Lambda $ = 0} \\
\hline
\underline {}electromagnetism&
the 4D4d quantized units of the 5$^{th}$ mass dimension \\
\hline
the strong force in pion&
the 4D4d quantized units of the 6$^{th}$ mass dimension \\
\hline
the weak interaction&
the 4D4d quantized units of the 7$^{th}$ mass dimension \\
\hline
\end{longtable}
}}

\bigskip

In the high-density region such as nucleus, the compression as the reverse
of the decomposition in Eq. (\ref{eq2}) takes place. The compression theory is the
reverse of the decomposition theory. In the compression theory, the D + 1
dimensional string in AdS$_{D+1}$ is the compression product of the CFT in
the form of the D dimensional particle and its virtual particle. The duality
in the compression theory is the CFT/AdS duality, the reverse of the AdS/CFT
duality. It can be expressed as string/particle duality.

One example is the big bang baryonenesis (the baryon formation) that
occurred immediately after the big bang. Within one minute after the big
bang before the big bang nucleosynthesis, when both density and energy were
high, baryons, such as proton and neutron, were formed from pions. Pion is
derived from the gauge force field as the 6$^{\rm th}$ mass dimension
(dimensional orbital) in binary lattice space (Chung, 2002a). The high
density compressed massive quark and massive pion as the strong force
(massive gluon) into the massive 5D string and tessellattice as AdS
$$
\qquad\qquad 1_{4} \,\pi \, = \,\,1_{4} \,1_{1} \ \ \ \  \buildrel {the\,\,\,compression{\kern
1pt} \,\,in\,\,\,nucleus} \over \longrightarrow \ \ \  \,{\begin{array}{*{20}c}
 {2_{5}}  \hfill & { + \,\,\,} \hfill \\
\end{array}} \,0_{5} \,\,\,\,\,\,\,\,\,\,\,\,\,\,\,\,\,\,\,\,\,\,\,\,\,\,
\eqno(24a)
$$

Eq. (24$a$) shows the CFT/AdS duality for the compression. The 5D string is
the string in the strong force as in the gauge particle/string duality
proposed by Polchinski and Strassler (2003). At low energy, quarks and
gluons are combined as the 5D strings. The different oscillations of this 5D
string produce different particles. The string can also explain some aspects
of masses and spins of the particles.

With the high-energy input after the big bang, the 5D string interacted with
AdS, resulting in the decomposition
$$
\begin{array}{l}
 \,\,{\begin{array}{*{20}c}
 {2_{5}}  \hfill & { + \,\,\,} \hfill \\
\end{array}} 0_{5} \ \ \ \  \buildrel {decomposition\,\,of\,space\,\,dimension}
\over \longrightarrow  \ \ \ \ \   1_{4} \,\,1_{1} \, = \,\,1_{4} \,i \\
 \,\,\,\,\,\, \\
 \end{array}
\eqno(24b)
$$
The result of the decomposition is 1$_{4}$1$_{1}$ that was the
quark-inerton. By carrying mass-energy of quark, inerton actually becomes
the major part of the constituent masses and the binding energy of quarks.
The inerton field is the ``auxiliary dimensional orbital'' as in Figure 3.
During the simultaneous slicing, the particles-antiparticles with the CP
symmetry between particle and antiparticle result in annihilation to become
cosmic radiation. The particles that are not annihilated have asymmetrical
charge-parity (CP asymmetry), in such way that the particle-antiparticle has
two asymmetrical sets of dimensional orbitals. The two sets of dimensional
orbitals are ``principal dimensional orbital'' and ``auxiliary dimensional
orbital''. The auxiliary set is dependent on the principal set, so the
particle-antiparticle appears to have only one set of dimensional orbital.
Auxiliary dimensional orbital is the hidden dimensional orbital.

For the four-mass-dimensional particle (baryonic matter), the two sets of
seven dimensional orbitals are principal dimensional orbital and auxiliary
dimensional orbital. These hierarchical dimensional orbitals are the force
and mass fields, including gravity, as shown in Figure 3.

\begin{figure}
\begin{center}
\includegraphics[scale=0.8]{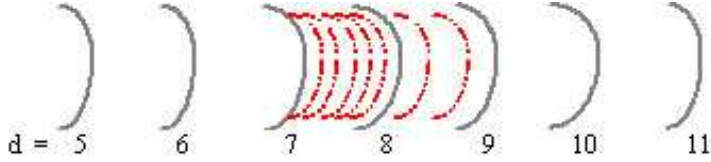}
\caption{\small{Dimensional orbitals in the four-mass dimensional (4d)
particle (baryonic matter): principal dimensional orbital (solid line), and
auxiliary dimensional orbital (dot line) as the inerton field; d is the
principal dimensional orbital number}}
\label{Figure 2}
\end{center}
\end{figure}

As shown in Figure 3, the seven orbitals of hidden auxiliary dimensional
orbital as the inerton field are in the middle of the seven orbitals of
principal dimensional orbital. The structure of the 4d particle with
dimensional orbitals resembles to the structure of atomic orbital.
Consequently, the periodic table of elementary particles is constructed to
account for all leptons, quarks, gauge bosons, and hadrons. The calculation
of the masses of quarks is the calculation of the masses in the inerton
field (see preliminary studies in (Chung, 2002a)), resulting in a good
agreement with the observable masses.

The requirement for the quark-inerton confinement confined 1$_{4}$
\textit{i}. The boundary of the confinement was dual to the boundary of AdS.
Thus, AdS was also in 1$_{4}$ \textit{i}. At high energy, the high velocity
of quark (1$_{4}$) resulted in the repetitive interactions between quark and
AdS to generate gluon, resulting in SU (\ref{eq3}). The space dimension of quark,
1$_{4}$, could not be decomposed further by AdS. Therefore, instead of the
decomposition of space dimension, the repetitive interactions between quark
(1$_{4}$) and AdS led to the decomposition of massive quark into nearly
massless quark, resulting in chiral symmetry of quark and asymptotic freedom
between quark and gluon at high energy
$$
\begin{array}{l}
 \,1_{4} \,i\, + \,\,AdS  \qquad\qquad\qquad   \buildrel {\,decomposition\,\,\,of\,mass} \over
\longrightarrow \  \\   \\
chiral\,\,symmetry\,\,and\,\,asymptoticly\,freedom\,\,for\,\,massless\,\,\,1_{4}
\,g \\
 \,\,\,\,\,\, \\
 \end{array}
\eqno(24c)
$$

The strong force in baryon is the combination of the CFT/AdS duality, Eq.
(24$a$), for the compression involving $\pi $, the AdS/CFT duality, Eq. (24$b$),
for the decomposition of space dimension involving inerton, and the
AdS/SU(3) duality, Eq. (24$c$), for the decomposition of mass involving gluon.
These three dualities during the baryonenesis form the base for the mass
calculation of hadrons (Chung, 2002a).

After the big bang baryonenesis, the universe was cool enough to form stable
protons and neutrons. The density was high enough for the big bang
nucleosynthesis, responsible for the formation of hydrogen, deuterium,
helium, and lithium. It lasted three minutes. Afterward, the temperature and
density of the universe fell below the condition for nuclear fusion.

\subsection*{7. Summary}

In the cosmic organism theory, different universes are the different cosmic
organs for the cosmic organism that is the multiverse. Different universes
are the different expressions of the common cosmic gene. Different stages in
a universe, like an organ, have different cosmic genetic expressions.

The cosmic gene is the cosmic digital code in the cosmic dimension. The
cosmic digital code includes the codes for the object structure and the
space structure. The cosmic digital code for the object structure consists
of full object (1, 2, and 3 for particle, string, and membrane,
respectively) and empty object (0) as AdS (anti de Sitter space). The
tessellation lattice of empty objects, which is constructed in the real
space, is tessellattice. The decomposition of a full object in tessellattice
results in the AdS/CFT duality. The digital code for the object structure
accounts for the AdS/CFT duality, the dS/bulk duality, and gravity.

The digital code for the space structure consists of 1 and 0 for attachment
space and detachment space, respectively. Attachment space attaches to
object permanently at zero speed or reversibly at the speed of light.
Detachment space detaches from the object irreversibly at the speed of
light. The combination of attachment space and detachment space results in
miscible space, binary lattice space or binary partition space. Miscible
space represents special relativity. Binary lattice space consists of
multiple quantized units of attachment space separated from one another by
detachment space. Binary lattice space corresponds to nilpotent universal
computational rewrite system (NUCRS) by Diaz and Rowlands. The gauge force
fields and wavefunction are in binary lattice space. Binary partition space
consists of separated continuous phases of attachment space and detachment
space. The collapse of wavefunction results in binary partition space. With
tessellattice and binary lattice space, 11D brane is reduced to 4D particle
surrounded by gravity and the gauge force fields.

The cosmic dimension is in the framework of varying dimension number (VDN).
In VDN, there are different speeds of light in different dimensional
space-times. In the VDN transformation, the decrease in the quantized speed
of light leads to the decrease in space-time dimension number (D) and the
increase of mass in terms of increasing mass dimension number (d). In the
same dimensional space-time, the varying supersymmetry transformation
carriers over mass in terms of mass dimension number.

The four stages in the maturation of our universe are the primitiverse, the
dual chiral universe, the achiral universe, and the dual Vedic universe.
Each stage has different cosmic genetic expressions. The most primitive part
of the multiverse is the primitiverse consisting of closely packed 10D
strings in attachment space. The primitiverse does not have detachment space
and empty object. The combination of two strings with additional space
dimension results in 11D brane and AdS as empty object. The interaction of
11D brane and empty object forms 10D string and pregravity through the
AdS/CFT duality. 10D strings and 10D antistrings are separated in the dual
chiral universe separated by the repulsion between pregravity and
anti-pregravity.

The collapse and the bounce of the dual chiral universe result in dual
achiral universe with 10D particles in both domains separated by the
repulsion between pregravity and anti-pregravity. The conversion from the
chiral dual universe to the achiral universe is predicted by the Santilli
isodual theory and the Dirac hole theory. With the different VDN
transformations and varying supersymmetry transformations, the two domains
become the light universe and the dark universe. The dark universe does not
have detachment space. The light universe has detachment space that slices
mass dimensions into 4D quantized units that are the origin of gauge force
fields. The combination of attachment space and detachment space is the
cosmic origin of quantum mechanics and gauge force fields. The hidden dark
universe develops into dark energy later. In the light universe, dark matter
has different mass numbers form baryonic matter.

The force fields in tessellattice for high-density region with boundary are
the original gravity before the big bang and the strong force in baryon. One
can correspond tessellattice to AdS space. The force fields in binary
lattice space for low-density region without boundary are electromagnetism,
the weak interaction, and the strong force in pion. Binary lattice space
corresponds to \textit{S}-matrix in gauge symmetry. The force fields in
tessellattice are the original gravity before the big bang, or a series of
small big bangs, which will be disclosed in a future paper under
preparation, and the strong force in baryon. The force fields in binary
lattice space are electromagnetism, the weak interaction, and the strong
force in pion.

\subsection*{References}

\noindent
Albrecht, A. and Steinhardt, P. J. (1982). ``Cosmology for Grand Unified
Theories with Radiatively Induced Symmetry Breaking'', Phys. Rev. Lett. 48,
pp. 1220-1223 .   \\

\noindent
Albrecht, A. and Magueijo, J. (1999). ``A time varying speed of light as a
solution to cosmological puzzles'', Phys. Rev. D59, 043516 (also
astro-ph/9811018). \\

\noindent
Amelino-Camelia, G. (2001). ``Testable scenario for Relativity with
minimum-length'', Phys. Letts. B510, pp. 255-263 (also hep-th/0012238).  \\

\noindent
Amelino-Camelia, G. (2002). ``Relativity in space-times with short-distance
structure governed by an observer-independent (Planckian) length scale'',
Int. J. Mod. Phys. D11, pp. 35-60 (also gr-qc/0012051).  \\

\noindent
Barrow, J. D. (1999). ``Cosmologies With Varying Light Speed'', Phys. Rev.
D59, 043515.  \\

\noindent
Barrow, J. D. (2003). ``Unusual Features of Varying Speed of Light
Cosmologies'', Phys.Lett. B564, pp. 1-7 (also gr-qc/0211074). \\

\noindent
Bell, J. S. (1964). ``On the Einstein-Podolsky-Rosen Paradox'', Physics 1,
pp. 195-199.  \\

\noindent
Bounias, M. (2002). ``On Spacetime Differential Elements and the
Distribution of Biohamiltonian Components'', Spacetime \& Substance 3, no.
1, pp. 15-19 (also physics/0205087).  \\

\noindent
Bounias, M. and Krasnoholovets, V. (2003a). ``Scanning the Structure of
Ill-known Spaces: Part 2. Principles of construction of physical space'',
Kybernetes: The Int. J. of Systems and Cybernetics 32, no. 7/8, pp. 976-1004
(also physics/0212004).  \\

\noindent
Bounias, M. and Krasnoholovets, V. (2003b). ``Scanning the Structure of
Ill-Known Spaces: Part 3. Distribution of Topological Structures at
Elementary and Cosmic Scales'', Kybernetes: The Int. J. Systems and
Cybernetics 32, no. 7/8, pp. 1005-1020 (also physics/0301049).   \\

\noindent
Bounias, M. and Krasnoholovets, V. (2003c). ``Scanning the Structure of
Ill-known Spaces: Part 1. Founding Principles About Mathematical
Constitution of Space'', Kybernetes: The Int. J. Systems and Cybernetics 32,
no. 7/8, pp. 945-975 (also physics/0211096).   \\

\noindent
Chung, D. (2002a). ``The Cosmic Organism Theory for the Multiverse'',
hep-th/0201115.   \\

\noindent
Chung, D. (2002b). ``The Cosmic Digital Code and Quantum Mechanics'',
quan-ph/ 0204033.  \\

\noindent
Diaz, B. M. and Rowlands, P. (2003). ``A Computational path to the Nilpotent
Dirac Equation'', Symposium 10, International Conference for Computing
Anticipatory Systems, HEC Liege, Belgium, August 11-16, 2003, American
Institute of Physics Proceedings of the International Conference of
Computing Anticipatory Systems, ed. Daniel Dubois.   \\

\noindent
Ellis, G., Kirchner, U. and Stoeger, W. (2004). ``Multiverses and Cosmology:
Philosophical Issues'', astro-ph/0407329.  \\

\noindent
Ellis, G. and Uzan, J. (2005). ```c' is the speed of light, isn't it?'', Am.
J. Phys. 73, pp. 240-247 (also gr-qc/0305099).  \\

\noindent
Gubser, S. S., Klebanov, I. R., and Polyakov, A. M. (1998) ``Gauge Theory
Correlators from Non-Critical String Theory'', Phys. Lett. B428, pp. 105-114
(also see hep-th/9802109).  \\

\noindent
Guth, A. H. (1981). ``The Inflationary Universe: A Possible Solution to the
Horizon and Flatness Problems'', Phys. Rev. D23, pp. 347-356.   \\

\noindent
Krasnoholovets, V. and Ivanovsky, D. (1993). ``Motion of a particle and the
vacuum'', Phys. Essays 6, no. 4, pp. 554-563 (also quant-ph/9910023).  \\

\noindent
Krasnoholovets, V. (1997). ``Motion of a relativistic particle and the
vacuum'', Phys. Essays 10, no. 3, pp. 407-416 (also quant-ph/9903077).  \\

\noindent
Krasnoholovets, V. (2000). ``On the nature of spin, inertia and gravity of a
moving canonical particle'', Indian Journal of Theoretical Physics
48, no. 2, pp. 97-132 (also quant-ph/0103110).   \\

\noindent
Krasnoholovets, V. (2002). ``Submicroscopic deterministic quantum
mechanics,'' Int. J. Computing Anticipatory Systems 11, pp. 164-179 (also
quant-ph/0109012).  \\

\noindent
Linde, A. D. (1982). ``New Inflationary Universe Scenario: A Possible
Solution Of The Horizon, Flatness, Homogeneity, Isotropy And Primordial
Monopole Problems'', Phys. Lett. B108, pp. 389-393.    \\

\noindent
Maldacena, J. (1998). ``The Large N Limit of Superconformal Field Theories
and Supergravity'', Adv. Theor. Math. Phys. 2, pp. 231-252, (also see
hep-th/9711200).   \\

\noindent
Magueijo, J. (2003). ``New varying speed of light theories'', Rep. Prog.
Phys. 66, pp. 2025-2068, (also astro-ph/0305457).  \\

\noindent
Penrose, R. (2000). ``Wavefunction Collapse as a Real Gravitational Effect''
in Mathematical Physics, eds: by A. Fokas, A. Grigoryan, T. Kibble \& B.
Zegarlinski (Imperial College, London), pp. 266-282.  \noindent

\noindent
Polchinski, J and Strassler, M. J. (2003) ``Deep Inelastic Scattering and
Gauge/String Duality'', J. High Energy Phys. 0305, 012 (also see
hep-th/0209211).  \\

\noindent
Roy, R. R. M., (1999), ``Vedic Physics: Scientific Origins of Hinduism'',
Golden Egg Publishing (Toronto, Canada).   \\

\noindent
Randall, L. and Sundrum, R. (1999a). ``Out of this World Supersymmetry
Breaking'', Nucl. Phys. B557, pp. 79-118. \\

\noindent
Rees, M. (2003). ``Dark Matter: Introduction'', Phil. Trans. Roy. Soc.
London 361, pp. 2427-2434.   \\

\noindent
Santilli, R. M., (2005), ``Isodual Theory Of Antimatter With Applications to
Antigravity, Grand Unification, and Cosmology'' (Kluwer Academic Publishers,
Boston/Dordrecht/London), in press.  \\

\noindent
Whitehead, A. (1929) ``Process and Reality'', (New York, Macmillan). \\

\noindent
Witten, K. (1998) ``Anti-de Sitter Space and Holography'', Adv. Theor. Math.
Phys. 2, pp. 253-291 (also see hep-th/9802150). \\

\end{document}